\documentclass[aps,preprint]{revtex4}%
\usepackage{amsfonts}
\usepackage{amsmath}
\usepackage{amssymb}
\usepackage{graphicx}%
\begin{document}
\preprint{Int. J. Appl. Math. Inf. Sci.}
\title[ ]{\textbf{Nucleon Properties from Approximating Chiral Quark Sigma Model}}
\author{\textbf{M. Abu-shady}}
\affiliation{Faculty of Science, Menoufia University, Shebin El-kom, Egypt. }
\author{}
\affiliation{}
\affiliation{}
\author{}
\affiliation{}

\begin{abstract}
{We apply the approximating chiral quark model.} This chiral quark model is
based on an effective Lagrangian which the interactions between quarks
via\ sigma and pions-mesons. {The field equations have been solved in the
mean-field approximation for the hedgehog baryon state. Good results are
obtained for nucleon properties in comparison with original model.}\newline

\end{abstract}
\volumeyear{ }
\volumenumber{ }
\eid{ }
\date{}
\received{March 2008}

\accepted{June 2008}

\published{}

\startpage{1}
\endpage{ }
\maketitle
\tableofcontents

\section{$\mathbf{Introduction}$}

The description of the processes involving strong interactions is very
difficult in the frame of the quantum chromodynamics (QCD) that is due to its
non-abelian color and flavor structure and strong coupling constants. These
effective models, like linear sigma model, are constructed in such a way as to
respect general properties from the more fundamental theory (QCD), such as the
chiral symmetry and its spontaneous breaking, e.g. [1, 2]. It is known that
the linear sigma model of Gell-Mann and Levy [1] does not always give the
correct phenomenology, e.g. the value of pion-nucleon sigma term is too large
as in Refs. [3, 4, 5]. Sigma term provides a direct measure of the scalar
quark condensates in the nucleon and thereby also constitute an indicator for
the mechanism of explicit chiral symmetry breaking. Birse and Banerjee [3]
constructed equations of motion treating both $\sigma$ and $\mathbf{\pi-}$
fields as time-independence classical fields and the quarks in hedgehog spinor
state that work is reexamined by Broniowski and Banerjee [4]. Birse [5]
generalized this mean-filed model to include angular momentum and isospin projections.

Recently, the mesons play a very important role in improving the nucleon
properties in the chiral quark models, the perturbative chiral quark model is
extended to include the kaon and eta-mesons cloud contributions, to analyze
the nucleon properties [6-9]. In the same direction, Horvat et al. [10]
applied Tamm-Dancoff method to the chiral quark model which is extended to
include additional degrees of freedom as a pseudoscalar-isoscalar field and a
triplet of scalar isovector to get better description of nucleon properties.
On the other hand, Broniowski and Golli [11] analyzed a particular extension
of the linear sigma model coupled to valence quarks, which contained an
additional term with gradients of the chiral fields and investigated the
dynamically consequence of this term and its relevant to the phenomenology of
the soliton models of the nucleon. In same direction, {Rashdan et al. [12, 13]
considered the higher-order mesonic interactions in the chiral quark sigma
model to get a better description of nucleon properties.} The aim of this
paper is to examine nucleon properties, such as sigma commutator, coupling
constant $g_{A}(0)$ and pion-nucleon coupling constant $g_{\pi NN}(0)$ which
are not calculated in Ref. [11].

The paper is organized as follows. Section II: We review briefly approximating
chiral quark sigma model. Numerical calculations and the results are presented
in section III.

\section{Approximating Quark Sigma Model}

Approximating chiral quark sigma model is described in details in Ref. [11].
The following we give a brief summary.

The Lagrangian density of approximating quark sigma model which describes the
interactions between quarks via\ the $\sigma-$and $\mathbf{\pi}-$mesons is
written as $\left[  11\right]  $
\begin{align}
L\left(  r\right)   &  =\overline{\Psi}i\gamma_{\mu}\partial^{\mu}\Psi
+\frac{1}{2}\left(  \partial_{\mu}\sigma\partial^{\mu}\sigma+\partial_{\mu
}\mathbf{\pi}.\partial^{\mu}\mathbf{\pi}\right)  +\frac{1}{2}A_{0}\left(
\sigma\partial^{\mu}\sigma+\mathbf{\pi}.\partial^{\mu}\mathbf{\pi}\right)
^{2}+g\overline{\Psi}\left(  \sigma+i\gamma_{5}\mathbf{\tau}.\mathbf{\pi
}\right)  \Psi\nonumber\\
&  -U_{1}\left(  \sigma,\mathbf{\pi}\right)  , \tag{2-1}%
\end{align}
with%

\begin{equation}
U_{1}\left(  \sigma,\mathbf{\pi}\right)  =\frac{\lambda_{1}^{2}}{4}\left(
\sigma^{2}+\mathbf{\pi}^{2}-\nu_{1}^{2}\right)  ^{2}+m_{\pi}^{2}f_{\pi}\sigma,
\tag{2-2}%
\end{equation}
is the meson-meson interaction potential where $\Psi,\sigma$ and $\mathbf{\pi
}$ are the quark, sigma and pion fields, respectively. In the mean-field
approximation the meson fields are treated as time-independent classical
fields. This means that we are replacing powers and products of the meson
fields by corresponding powers and products of their expectation values. The
meson-meson interactions in Eq.(2-2) leads to hidden chiral $SU(2)\times
SU(2)$ symmetry with $\sigma\left(  r\right)  $ taking on a vacuum expectation
value \ \
\begin{equation}
\ \ \ \ \ \ \left\langle \sigma\right\rangle =-f_{\pi}, \tag{2-3}%
\end{equation}
where $f_{\pi}=93$ MeV is the pion decay constant. The final \ term in Eq.
(2-2) is included to break the chiral symmetry. It leads to partial
conservation of axial-vector isospin current (PCAC). In the original model
[3], the A-term is excluded by the requirement of renormalizability. Since we
are going to use Eq. (2-1) as an effective model, approximating the underlying
quark theory, the model need not and should not be renormalizable as in Refs.
[10-13]. The parameters $\lambda_{1}^{2},$ $\nu_{1}^{2}$ are expressed in
terms of $f_{\pi}$ and the masses $\sigma-$and $\mathbf{\pi}- $mesons, we get
\begin{equation}
\lambda_{1}^{2}=\frac{\bar{m}_{\sigma}^{2}-m_{\pi}^{2}}{4f_{\pi}^{2}},
\tag{2-4}%
\end{equation}%
\begin{equation}
\nu_{1}^{2}=f_{\pi}^{2}-\frac{m_{\pi}^{2}}{\lambda_{1}^{2}}, \tag{2-5}%
\end{equation}%
\begin{equation}
\bar{m}_{\sigma}^{2}=\left(  1+f_{\pi}^{2}A_{0}\right)  m_{\sigma}^{2}
\tag{2-6}%
\end{equation}
(for details, see Ref. [11])

Now we expand the extremum, with the shifted field defined as
\begin{equation}
\sigma=\sigma^{\prime}-f_{\pi}, \tag{2-7}%
\end{equation}
substituting by Eq. (2-7) into Eq. (2-1), we get%

\begin{align}
L\left(  r\right)   &  =\overline{\Psi}i\gamma_{\mu}\partial^{\mu}\Psi
+\frac{1}{2}\left(  \partial_{\mu}\sigma^{\prime}\partial^{\mu}\sigma^{\prime
}+\partial_{\mu}\mathbf{\pi}.\partial^{\mu}\mathbf{\pi}\right)  +\frac{1}%
{2}A_{0}\left(  \sigma^{\prime}\partial^{\mu}\sigma^{\prime}+\mathbf{\pi
}.\partial^{\mu}\mathbf{\pi}\right)  ^{2}-g\overline{\Psi}f_{\pi}%
\Psi+\nonumber\\
&  g\overline{\Psi}\sigma^{\prime}\Psi+ig\overline{\Psi}\mathbf{\gamma}%
_{5}.\mathbf{\pi}\Psi-U_{1}\left(  \sigma^{\prime},\mathbf{\pi}\right)
\tag{2-8}%
\end{align}
with
\begin{equation}
U_{1}\left(  \sigma^{\prime},\mathbf{\pi}\right)  =\frac{\lambda_{1}^{2}}%
{4}(\left(  \sigma^{\prime}-f_{\pi})^{2}+\mathbf{\pi}^{2}-\nu_{1}^{2}\right)
^{2}+m_{\pi}^{2}f_{\pi}(\sigma^{\prime}-f_{\pi}).\nonumber
\end{equation}

The time-independent fields $\sigma^{^{\prime}}\left(  r\right)  \,\,$and
$\mathbf{\pi}\left(  r\right)  $ satisfy the Euler-Lagrange equation, and the
quark wave function satisfies the Dirac eigenvalue equation. Substituting by
Eq. (2-8) in Euler-Lagrange equation as in Ref. [11], we obtain%

\begin{align}
\square\sigma^{\prime}  &  =\frac{-1}{\left(  1+((\sigma^{\prime}-f_{\pi}%
)^{2}+\mathbf{\pi}^{2})A_{0}\right)  }\{(1+\mathbf{\pi}^{2}A_{0}%
)((A_{0}(\sigma^{\prime}-f_{\pi})(\left(  \partial^{\mu}\sigma^{\prime
}\right)  ^{2}+\left(  \partial^{\mu}\mathbf{\pi}\right)  ^{2})+\frac{\partial
U_{1}}{\partial\sigma^{\prime}}-g\overline{\Psi}\Psi)\nonumber\\
&  -A_{0}(\sigma^{\prime}-f_{\pi})\mathbf{\pi(}(A_{0}\mathbf{\pi}\left(
\left(  \partial^{\mu}\sigma^{\prime}\right)  ^{2}+\left(  \partial^{\mu
}\mathbf{\pi}\right)  ^{2}\right)  +\frac{\partial U_{1}}{\partial\mathbf{\pi
}}-ig\overline{\Psi}\gamma_{5\cdot}\mathbf{\tau}\Psi)\} \tag{2-9}%
\end{align}%
\begin{align}
\square\mathbf{\pi}  &  =\frac{-1}{\left(  1+((\sigma^{\prime}-f_{\pi}%
)^{2}+\mathbf{\pi}^{2})A_{0}\right)  }\{(-A_{0}(\sigma^{\prime}-f_{\pi
})\mathbf{\pi)((}A_{0}(\sigma^{\prime}-f_{\pi})(\left(  \partial^{\mu}%
\sigma^{\prime}\right)  ^{2}+\left(  \partial^{\mu}\mathbf{\pi}\right)
^{2})+\frac{\partial U_{1}}{\partial\sigma^{\prime}}-\nonumber\\
&  g\overline{\Psi}\Psi)+(1+(\sigma^{\prime}-f_{\pi})^{2}A_{0})(A_{0}%
\mathbf{\pi(}\left(  \partial^{\mu}\sigma^{\prime}\right)  ^{2}+\left(
\partial^{\mu}\mathbf{\pi}\right)  ^{2})+\frac{\partial U_{1}}{\partial
\mathbf{\pi}}-ig\overline{\Psi}\gamma_{5\cdot}\mathbf{\tau}\Psi)\} \tag{2-10}%
\end{align}
where $\mathbf{\tau}$ refers to Pauli isospin matrices and $\gamma_{5}=\left(
\begin{array}
[c]{cc}%
0 & 1\\
1 & 0
\end{array}
\right)  .$ If $A_{0}=0,$ the usual Birse and Banerjee [3] model equations of
motion are recovered. Including the color degree of freedom, one has
$g\overline{\Psi}\Psi\rightarrow N_{c}g\overline{\Psi}\Psi$ where $N_{c}=3$
colors. Thus
\begin{equation}
\Psi\left(  r\right)  =\frac{1}{\sqrt{4\pi}}\left[
\begin{array}
[c]{c}%
u\left(  r\right) \\
iw\left(  r\right)
\end{array}
\right]  \qquad\text{and}\qquad\bar{\Psi}\left(  r\right)  =\frac{1}%
{\sqrt{4\pi}}\left[
\begin{array}
[c]{cc}%
u\left(  r\right)  & iw\left(  r\right)
\end{array}
\right]  , \tag{2-11}%
\end{equation}
and the sigma, pion and vector densities are given by
\begin{align}
\rho_{s}  &  =N_{c}\overline{\Psi}\Psi=\frac{3}{4\pi}\left(  u^{2}%
-w^{2}\right)  ,\tag{2-12}\\
\rho_{p}  &  =iN_{c}\overline{\Psi}\gamma_{5}\vec{\tau}\Psi=\frac{3}{4\pi
}g\left(  -2uw\right)  ,\tag{2-13}\\
\rho_{v}  &  =\frac{3}{4\pi}\left(  u^{2}+w^{2}\right)  , \tag{2-14}%
\end{align}

The boundary conditions for the asymptotics for $\sigma\left(  r\right)  $ and
$\pi\left(  r\right)  $ at $r\rightarrow\infty$ are:
\begin{equation}
\sigma\left(  r\right)  \sim-f_{\pi}\text{,}\qquad\pi\left(  r\right)
\sim0\text{ } \tag{2-15}%
\end{equation}
We used the hedgehog ansatz [3], where
\begin{equation}
\mathbf{\pi}\left(  r\right)  =\overset{\symbol{94}}{\mathbf{r}}\pi\left(
r\right)  . \tag{2-16}%
\end{equation}
The chiral Dirac equation for the quarks is [12]
\begin{equation}
\frac{du}{dr}=-P\left(  r\right)  u+\left(  W+m_{q}-S(r)\right)  w, \tag{2-17}%
\end{equation}
where the scalar potential $S(r)=g\left\langle \sigma^{\prime}\right\rangle
,$the pseudoscalar potential $P(r)=\left\langle \mathbf{\pi}.\hat
{r}\right\rangle $, and $W$ is the eigenvalue of the quarks spinor $\Psi.$%
\begin{equation}
\frac{dw}{dr}=-\left(  W-m_{q}+S(r)\right)  u-\left(  \frac{2}{r}-P\left(
r\right)  \right)  w. \tag{2-18}%
\end{equation}

\section{Numerical Calculations}

\subsection{The scalar field $\sigma^{\prime}$}

To solve Eq. (2-9), we integrate a suitable Green's function over the source
fields as in Ref. $\left[  14\right]  .$\ \ Thus
\begin{align}
\sigma^{\prime}\left(  \mathbf{r}\right)   &  =\int d^{3}\mathbf{r}^{\prime
}D_{\sigma}(\mathbf{r-\grave{r}})[\frac{-1}{\left(  1+((\sigma^{\prime}%
-f_{\pi})^{2}+\mathbf{\pi}^{2})A_{0}\right)  }\{(1+\mathbf{\pi}^{2}%
A_{0})(A_{0}(\sigma^{\prime}-f_{\pi})(\left(  \partial^{\mu}\sigma^{\prime
}\right)  ^{2}\nonumber\\
&  +\left(  \partial^{\mu}\mathbf{\pi}\right)  ^{2})+\frac{\partial U_{1}%
}{\partial\sigma^{\prime}}-g\rho_{s}(\mathbf{\grave{r}}))-A_{0}(\sigma
^{\prime}-f_{\pi})\mathbf{\pi(}A_{0}\mathbf{\pi}\left(  \left(  \partial^{\mu
}\sigma^{\prime}\right)  ^{2}+\left(  \partial^{\mu}\mathbf{\pi}\right)
^{2}\right)  +\nonumber\\
&  \frac{\partial U_{1}}{\partial\mathbf{\pi}}-g\rho_{p}(\mathbf{\grave{r}%
}))\}] \tag{3-19}%
\end{align}
where
\[
D_{\sigma}(\mathbf{r-\grave{r}})=\frac{1}{4\pi\left\vert \mathbf{r-\grave{r}%
}\right\vert }\exp(-m_{\sigma}\left\vert \mathbf{r-\grave{r}}\right\vert ),\;
\]
the scalar field is spherical in this model as we only need the $l=0$\ term
\begin{equation}
D_{\sigma}\left(  \mathbf{r-\grave{r}}\right)  =\frac{1}{4\pi}\sinh\left(
m_{\sigma}r_{<}\right)  \frac{\exp\left(  -m_{\sigma}r_{>}\right)  }{r_{>}%
},\;\; \tag{3-20}%
\end{equation}
therefore we arrive at the integral equation for $\sigma^{\prime}\left(
\mathbf{r}\right)  :$
\begin{align}
\sigma^{\prime}\left(  \mathbf{r}\right)   &  =m_{\sigma}\int\limits_{0}%
^{\infty}r^{\prime2}dr^{\prime}(\frac{\sinh\left(  m_{\sigma}r_{>}\right)
}{m_{\sigma}r_{>}}\frac{\exp\left(  -m_{\sigma}r_{>}\right)  }{m_{\sigma}%
r_{>}})[\frac{-1}{\left(  1+((\sigma^{\prime}-f_{\pi})^{2}+\mathbf{\pi}%
^{2})A_{0}\right)  }\{(1+\mathbf{\pi}^{2}A_{0})\times\nonumber\\
&  \times(A_{0}(\sigma^{\prime}-f_{\pi})(\left(  \partial^{\mu}\sigma^{\prime
}\right)  ^{2}+\left(  \partial^{\mu}\mathbf{\pi}\right)  ^{2})+\frac{\partial
U_{1}}{\partial\sigma^{\prime}}-g\rho_{s}(\mathbf{\grave{r}}))-\nonumber\\
&  A_{0}(\sigma^{\prime}-f_{\pi})\mathbf{\pi(}A_{0}\mathbf{\pi}\left(  \left(
\partial^{\mu}\sigma^{\prime}\right)  ^{2}+\left(  \partial^{\mu}\mathbf{\pi
}\right)  ^{2}\right)  +\frac{\partial U_{1}}{\partial\mathbf{\pi}}-g\rho
_{p}(\mathbf{\grave{r}}))\}] \tag{3-21}%
\end{align}
We will solve these implicit integral equation by iterating to self consistency.

\subsection{The pion field $\mathbf{\pi}$}

To solve Eq. (2-10) we integrate a suitable Green's function over the source
fields. We use $l=1$\ component of the pion Green's function. Thus
\begin{align}
\;\;\;\ \ \mathbf{\pi}\left(  r\right)   &  =m_{\pi}\int_{0}^{\infty}%
r^{\prime2}dr^{\prime}\frac{[-\sinh\left(  m_{\pi}r_{<}\right)  +m_{\pi}%
r_{<}\cosh\left(  m_{\pi}r_{<}\right)  ]}{\left(  m_{\pi}r_{>}\right)  ^{2}%
}\times
\;\;\;\;\;\;\;\;\;\;\;\;\;\;\;\;\;\;\;\;\;\;\;\;\;\;\;\;\;\;\;\;\;\nonumber\\
&  \;\times\frac{-1}{\left(  1+((\sigma^{\prime}-f_{\pi})^{2}+\mathbf{\pi}%
^{2})A_{0}\right)  }\{-A_{0}(\sigma^{\prime}-f_{\pi})(\mathbf{\pi}A_{0}%
(\sigma^{\prime}-f_{\pi})(\left(  \partial^{\mu}\sigma^{\prime}\right)
^{2}+\left(  \partial^{\mu}\mathbf{\pi}\right)  ^{2})+\nonumber\\
&  \frac{\partial U_{1}}{\partial\sigma^{\prime}}-g\rho_{s}(\mathbf{\grave{r}%
}))+(1+(\sigma^{\prime}-f_{\pi})^{2}A_{0})(A_{0}\mathbf{\pi(}\left(
\partial^{\mu}\sigma^{\prime}\right)  ^{2}+\left(  \partial^{\mu}\mathbf{\pi
}\right)  ^{2})+\nonumber\\
&  \frac{\partial U_{1}}{\partial\mathbf{\pi}}-g\rho_{p}(\mathbf{\grave{r}%
})))\} \tag{3-22}%
\end{align}
We have solved Dirac Eqs. (2-17, 2-18) using fourth order Rung-Kutta method.
Due to the implicit nonlinearly of these Eqs. (2-9, 2-10) it is necessary to
iterate the solution until self-consistency is achieved. To start this
iteration process we use the chiral circle form for the meson fields:
\begin{equation}
S(r)=m_{q}(1-\cos\theta)\text{ and }P(r)=-m_{q}\sin\theta, \tag{3-23}%
\end{equation}
\ \ \ \ \ \ \ \ \ \ \ \ \ \ where $\theta=\tanh r$

\subsection{Properties of the Nucleon}

The proton and neutron magnetic moments are given by
\begin{equation}
\mu_{p,n}=<P\uparrow\left\vert \int d^{3}\mathbf{r}\frac{1}{2}\mathbf{r}%
\times\mathbf{j}_{\varepsilon M}(\mathbf{r})\right\vert P\uparrow>, \tag{3-24}%
\end{equation}

where, the electromagnetic current is
\begin{equation}
j_{\epsilon M}(\mathbf{r})=\bar{\Psi}\left(  \mathbf{r}\right)  \mathbf{\gamma
}\left(  \frac{1}{6}+\frac{\tau_{3}}{2}\right)  \Psi(\mathbf{r})-\varepsilon
_{\alpha\beta_{3}}\pi_{\alpha}\left(  \mathbf{r}\right)  \mathbf{\nabla}%
\pi_{\beta}\left(  \mathbf{r}\right)  , \tag{3-25}%
\end{equation}

such that
\begin{equation}
\left(  \mathbf{j}_{\epsilon M}(\mathbf{r})\right)  _{nucleon}=\bar{\Psi
}\left(  \mathbf{r}\right)  \mathbf{\gamma}\left(  \frac{1}{6}+\frac{\tau_{3}%
}{2}\right)  \Psi\left(  \mathbf{r}\right)  , \tag{3-26}%
\end{equation}%
\begin{equation}
\left(  \mathbf{j}_{\epsilon M}(\mathbf{r})\right)  _{meson}=-\epsilon
_{\alpha\beta3}\pi_{\alpha}\left(  \mathbf{r}\right)  \mathbf{\nabla}%
\pi_{\beta}\left(  \mathbf{r}\right)  , \tag{3-27}%
\end{equation}

\bigskip The nucleon axial-vector coupling constant is found from%
\begin{equation}
\frac{1}{2}g_{A}(0)=\left\langle P\uparrow\left\vert \int d^{3}rA_{3}%
^{z}(\mathbf{r})\right\vert P\uparrow\right\rangle , \tag{3-28}%
\end{equation}

where the z-component of the axial vector current is given by
\begin{equation}
A_{3}^{z}(\mathbf{r})=\bar{\Psi}\left(  \mathbf{r}\right)  \frac{1}{2}%
\gamma_{5}\gamma^{3}\tau_{3}\Psi\left(  \mathbf{r}\right)  -\sigma\left(
\mathbf{r}\right)  \frac{\partial}{\partial z}\pi_{3}\left(  \mathbf{r}%
\right)  +\pi_{3}\left(  \mathbf{r}\right)  \frac{\partial}{\partial z}%
\sigma\left(  \mathbf{r}\right)  . \tag{3-29}%
\end{equation}
The pion-nucleus $\sigma$ commutator is defined
\begin{equation}
\sigma(\pi N)=\left\langle P\uparrow\left\vert \int d^{3}r\sigma^{\prime
}(\mathbf{r})\right\vert P\uparrow\right\rangle , \tag{3-30}%
\end{equation}
In calculation of $\sigma(\pi N),$ we replace $\sigma^{\prime}(\mathbf{r}) $
by $\frac{j_{\sigma}(\mathbf{r})}{m_{\sigma}^{2}}$ where $j_{\sigma
}(\mathbf{r})$ is the source current defined by%

\[
(\square+m_{\sigma}^{2})\sigma^{\prime}=j_{\sigma}(\mathbf{r})
\]
\quad To calculate the pion-nucleon coupling constant, we consider the limit
$\mathbf{q}\longrightarrow0$ of%
\begin{equation}
\frac{g_{\pi NN}(0)}{2M}=<P\uparrow\left\vert \int d^{3}\mathbf{r}%
e^{i\mathbf{q}\cdot\mathbf{r}}\times\mathbf{j}_{\pi3}(\mathbf{r})\right\vert
P\uparrow>,\quad\tag{3-31}%
\end{equation}
where pion source current is defined by%

\begin{equation}
\bigskip(\square+m_{\mathbf{\pi}}^{2})=\mathbf{j}_{\pi3}(\mathbf{r})
\tag{3-32}%
\end{equation}

(For details see Refs. [3-5]).

\ \ \ \ \ \ \ \ \ \ \ \ \ \ \ \ \ \ \ \ \ \ \ \ \ \ \ \ \ \ \ \ \ \ \ \ \ \ \ \ \ \ \ \ \ \ \ \ \ \ \ \ \ \ \ \ \ \ \ \ \ \ \ \ \ \ \ \ \ \ \ \ \ \ \ \ \ \ \ \ \ \ \ \ \ \ \ \ \ \ \ \ \ \ \ \ \ \ \ \ \ \ \ 

\subsection{Discussion of Results}

The field equations (2.9$\rightarrow2.$18) have been solved by iteration as in
Ref. [10] for different values of quark and sigma masses. Tables 1, 2 and 3
show the nucleon observables calculated for $m_{q}=$400 - 480 MeV and
$m_{\sigma}=441$ MeV and $m_{\sigma}=900$ MeV, respectively.

From Table I, \ the nucleon properties are calculated at $m_{\sigma}=441$ MeV
which is predicted by Chiral Perturbation Theory (ChPT) [15]. Note, magnetic
moments of the nucleon are improved by increasing the quark mass that backs to
the increases of mesonic interactions with quarks. Also, Sigma commutator
$\sigma\left(  \pi N\right)  $ is improved by decreasing the quark mass. The
similar situation is hold for pion-nucleon coupling constant ($g_{\pi
NN}(0)\frac{m_{\pi}}{2M_{N}})$ which improved by decreasing the quark mass,
$g_{A}(0)$ is not little sensitive for changing in quark mass which backs to
the expression of axial vector current (Eq. 3-29) is not depend on explicitly
of quark mass but only on the dynamic of fields.

From Table II, the nucleon properties are calculated at $m_{\sigma}=900$ MeV
as in Ref [11]. We obtain good \ results at $m_{q}=$462 MeV which is
consistent with NJL model as in Ref. [16]. From Tables (I, II), we note the
$g_{A}(0)$ is little sensitive for sigma mass that backs to the same reason as
pointed out before. Also, the properties of the nucleon are not sensitive for
sigma mass [11], so we take two extreme values $m_{\sigma}=441$MeV and
$m_{\sigma}=900$ MeV which is compatible with Refs. [11, 15].

We known sigma commutator is important quantity to measure the breaking chiral
symmetry and it is one of problems in the full chiral sigma model as in Refs.
[3, 4, 5]. Recently, Sigma commutator is predicted by perturbative chiral
quark model [7] and lattice QCD [17] which the value obtained lies in the
range 45 to 55 MeV. The effect A-term is strongly on this quantity that the
change in range 30\% relative to full model [3]. Good value obtained equal 56
MeV at ($m_{q}=400$ and $m_{\sigma}=900)$MeV (see, Table II)

From Fig. 1, we see the sigma field passes through zero at $r=0.5$, which we
will refer to as the soliton radius $(r=0.5)$, whereas at $r\rightarrow\infty$
the pion field $\rightarrow0$ and sigma field $\rightarrow-f_{\pi}.$ The pion
field takes the shape of the P-wave, which gives the attraction of the
pion-quark interaction, and goes to zero in a linear manner for large
distances. We also see that the meson fields do not stray far from the
circular minimum of the potential $\sigma^{2}+\mathbf{\pi}^{2}=f_{\pi}^{2} $.
The pion field reaches its maximum value close to the soliton radius. Fig. II.
shows the components of quarks $u(r)$ and $w(r)$ corresponding the fields.

\textbf{Table I.} Values of magnetic moments of the nucleon{\small ,}
$\sigma(\pi N)$ term$,$ $g_{A}(0)$ and $g_{\pi NN}(0)\frac{m_{\pi}}{2M_{N}}$.
At $m_{6}=441$ MeV. $f_{\pi}^{2}A_{0}=-0.02.$ All quantities in MeV.%

\begin{tabular}
[c]{|l|l|l|l|l|l|}\hline
{\small \ }$m_{q}\left(  \text{MeV}\right)  $ & 400 & 420 & 440 & 462 &
480\\\hline
$\mu_{p}\left(  N\right)  $ & 2.572 & 2.640 & 2.699 & 2.757 & 2.845\\\hline
$\mu_{n}\left(  N\right)  $ & -1.873 & -1.951 & -2.018 & -2.085 &
-2.135\\\hline
$\sigma\left(  \pi N\right)  $ & 69 & 76 & 81 & 84 & 85\\\hline
$g_{A}(0)$ & 1.689 & 1.714 & 1.734 & 1.752 & 1.764\\\hline
$g_{\pi NN}(0)\frac{m_{\pi}}{2M_{N}}$ & 1.315 & 1.365 & 1.410 & 1.454 &
1.485\\\hline
\end{tabular}

\bigskip

\textbf{Table II.} Values of magnetic moments of the nucleon, $\sigma(\pi N)$
term$,$ $g_{A}(0)$ and $g_{\pi NN}(0)\frac{m_{\pi}}{2M_{N}}$. At $m_{6}=900$
MeV., $f_{\pi}^{2}A_{0}=-0.03.$ All quantities in MeV.%

\begin{tabular}
[c]{|l|l|l|l|l|l|}\hline
{\small \ }$m_{q}\left(  \text{MeV}\right)  $ & 400 & 420 & 440 & 462 &
480\\\hline
$\mu_{p}\left(  N\right)  $ & 2.509 & 2.581 & 2.632 & 2.676 & 2.706\\\hline
$\mu_{n}\left(  N\right)  $ & -1.908 & -1.983 & -2.039 & -2.087 &
-2.120\\\hline
$\sigma\left(  \pi N\right)  $ & 56 & 62 & 66 & 69 & 71\\\hline
$g_{A}(0)$ & 1.733 & 1.757 & 1.774 & 1.787 & 1.794\\\hline
$g_{\pi NN}(0)\frac{m_{\pi}}{2M_{N}}$ & 1.27 & 1.318 & 1.354 & 1.388 &
1.417\\\hline
\end{tabular}

\bigskip

\section{Comparison With Other Models}

It is interesting to compare the nucleon properties in the present approach
with other models. Here we consider two models: Perturbative Chiral Quark
Model [6-9] and Original sigma model [3]. The perturbative chiral quark model
is an effective model of baryons based on chiral symmetry. The baryon is
described as a state of three localized relativistic quarks supplemented by a
pseudoscalar meson cloud as dictated by chiral symmetry requirements. In this
model the effect of the meson cloud is evaluated perturbatively in a
systematic fashion. The model has been successfully applied to the nucleon
properties (see Table III). We obtained reasonable results in comparison with
this model which backs to perturbative chiral quark model based on non-linear
$\sigma-$ model Lagrangian and leads to good description of nucleon properties
(for details, see Refs. [6-9]). In particular, nucleon magnetic moments are
improved in comparison with this model. In comparison with full model of Birse
and Banerjee [3]. We note the most observables are sensitive for A-term that
best results are obtained at ($m_{q}=462$ and $m_{\sigma}=900)$MeV

\textbf{Table III}. Observables of the nucleon comparison with Quark Sigma
Model [3] and Perturbative Chiral Quark Model [6-9]%

\begin{tabular}
[c]{|l|l|l|l|l|}\hline
Quantity &
\begin{tabular}
[c]{l}%
$m_{\sigma}=900$ MeV\\
$m_{q}=462$ MeV
\end{tabular}
& [3] & [6-9] & Exp.\\\hline
$\mu_{p}\left(  N\right)  $ & 2.68 & 2.87 & 2.62$\pm0.02$ & 2.79\\\hline
$\mu_{n}\left(  N\right)  $ & -2.08 & --2.29 & -2.0$\pm0.02$ & -1.91\\\hline
$\sigma(\pi N)$ & 69 & 92 & 54.7 & -\\\hline
$g_{A}(0)$ & 1.78 & 1.86 & 1.19 & 1.25\\\hline
$g_{\pi NN}(0)\frac{m_{\pi}}{2M_{N}}$ & 1.38 & 1.53 & - & 1.0\\\hline
\end{tabular}

\bigskip%

{\parbox[b]{4.8871in}{\begin{center}
\includegraphics[
natheight=6.661700in,
natwidth=7.195200in,
height=4.5316in,
width=4.8871in
]%
{KUUJY002.wmf}%
\\
{\small Fig. 1: Sigma and pion fields (in units of }$f_{\pi}${\small ) as
functions in distance R for }${\small m}_{\sigma}=900$ MeV, {\small \ }%
${\small m}_{q}=462$ MeV, $f_{\pi}^{2}A_{0}=-0.03$%
\end{center}}}
%

{\parbox[b]{5.054in}{\begin{center}
\includegraphics[
natheight=6.661700in,
natwidth=7.195200in,
height=4.6855in,
width=5.054in
]%
{KUUJY003.wmf}%
\\
{\small Fig. 2: The components of quark (in units of }$f_{\pi})${\small \ as
function in distance R for }$m_{\sigma}=900${\small \ MeV, }$m_{q}%
=462${\small \ MeV, }$f_{\pi}^{2}A_{0}=-0.03$%
\end{center}}}

\section{Conclusion}

From the results, the most nucleon properties are improved in comparison with
original model. In particular, sigma commutator is improved in range 30\% in
comparison with full model. $g_{A}(0)$ is little sensitive for A-term, so we
need to test the quantum effects or increase mesonic interactions in this
approach to improve this quantity in future works.

\section{Acknowledgments}

The work was supported by the government of Egypt and the author thanks Prof.
Faessler for hosting to Institute of Theoretical Physics.\newpage

\section{$\mathbf{\operatorname{Re}fernces}$}

\begin{enumerate}
\item \textbf{M. Gell-Mann, M. Levy}, The Axial Vector Current in Beta Decay,
Nuovo Cimento \textbf{16}, 705 (1960).

\item \textbf{S. Gasiorowicz and D. A. Geffen}, Effective Lagrangians and
field algebras with chiral symmetry, Rev. Mod. Phys. V. \textbf{41}, N. 531, (1969).

\item \textbf{M. Birse and M. Banerjee}, Chiral model for nucleon and delta,
Phys. Rev. D\textbf{31}, 118 (1985).

\item \textbf{W. Broniowski and M. K. Banerjee},Chiral model of N and $\Delta$
with vector mesons, Phys. Lett. \textbf{158}B, 335, (1985).

\item \textbf{M. Birse}, Chiral model of the nucleon: Projecting the hedgehog
as a coherent state, Phys. Rev. D33, 1934 (1986).

\item \textbf{T. Inoue, V.~E.~Lyubovitskij, T.~Gutsche and A.~Faessle}r,
Updated analysis of meson-nucleon sigma terms in the perturbative chiral qurak
model, Phys. Rev. C69, 035207 (2004).

\item \textbf{V.~E.~Lyubovitskij, T.~Gutsche and A.~Faessler}, Nucleon
Properties in the Perturbative Chiral Quark Model, Phys.\ Rev.\ C\textbf{\ }%
64\textbf{,} 065203 (2001).

\item \textbf{K. Khosonthongkee, V. E. Lyubovitskij, Th. Gutsche, Amand
Faessler, K. Pumsa-ard, S. Cheedket\ and Y. Yan}, Axial form factor of the
nucleon in the perturbative chiral quark model, J. Phys. G\textbf{30}, (2004). PP.793-810.

\item \textbf{T. Inoue, V. E. Lyubovitskij, Th. Gutsche and Amand Faessler},
Ground-State Baryon Masses in the Perturbative Chiral Quark Model, Int. J.
Mod. Phys. E\textbf{15}, (2006)PP. 121-14

\item \textbf{D. Horvat, D. Horvatic, B. Podobnik and D. Tadic}, The extended
chiral qurak model in a Tamm- Dancoff inspired approximation, FIZIKA\ B,
\textbf{9}, (2000) 4, PP.181-196.

\item \textbf{W. Broniowski and B. Golli}, Approximating chiral quark models
with linear sigma models, Nucl. Phys. A \textbf{714,} (2003), PP.575-588.

\item \textbf{M. Rashdan, M. Abu-Shady and T.S.T. Ali}, Nucleon properties
form modified sigma model, Int. J. Mod. Phys. A, V. \textbf{22}, Nos.14 \&15
(2007), PP.2673-2681.

\item \textbf{M. Abu-Shady, M. Rashdan and T.S.T. Ali,} $N\pi$\textbf{\ }%
Scattering and Electromagnetic Corrections in the Extended Linear Sigma
Model\textbf{, }FIZAKA\ B (Zagreb) \textbf{16 }(2007) 1, PP.59-66

\item \textbf{E. K. Steven and C. M. Dawn; S. E. Koonin and D. C. Meredith},
Computational Physics (Fortran version), Addison Wesley Publishing Company (1990).

\item \textbf{H. Leutwyler}, Mass and width of the sigma, Int. Mod. Phys.
A\textbf{22}, (2007) PP. 257-265.

\item \textbf{M. Rashdan}, NN interaction derived from the Nambu--Jona-Lasinio
model, Chaos, Solitons Fractals \textbf{18}, 107, (2003).

\item \textbf{D. B. Leinweber, A.W. Thomas and S. V. Wright}, Lattice QCD
calculation of the sigma commutator, Phys. Lett. B \textbf{482} (2000), PP.109--113
\end{enumerate}

\end{document}